\documentclass[12pt]{article}
\usepackage{amsmath}
\usepackage{amsfonts}
\usepackage{epsfig}
\usepackage{graphicx}

\begin{document}

\title{Non-commutative tomography: A tool for data analysis and signal
processing}
\author{F. Briolle\thanks{%
Centre de Physique Th\'{e}orique, CNRS Luminy, case 907, F-13288 Marseille
Cedex 9, France, Francoise.Briolle@univmed.fr}, V. I. Man'ko\thanks{%
P.N.~Lebedev Physical Institute, Moscow, Russia, manko@sci.lebedev.ru}, B.
Ricaud\footnotemark[1] and R. Vilela Mendes\thanks{%
CMAF, Complexo Interdisciplinar, Universidade de Lisboa, Av. Gama Pinto, 2 -
P1699 Lisboa Codex, Portugal, e-mail: vilela@cii.fc.ul.pt,
http://label2.ist.utl.pt/vilela/} \thanks{%
IPFN, Instituto Superior T\'{e}cnico, Av. Rovisco Pais, Lisboa, Portugal} }
\date{ }
\maketitle

\begin{abstract}
Tomograms, a generalization of the Radon transform to arbitrary pairs of
non-commuting operators, are positive bilinear transforms with a rigorous
probabilistic interpretation which provide a full characterization of the
signal and are robust in the presence of noise. We provide an explicit
construction of tomogram transforms for many pairs of noncommuting operators
in one and two dimensions and illustrations of their use for denoising,
detection of small signals and component separation.
\end{abstract}

\section{Introduction}

Integral transforms \cite{handbook} \cite{BWolf79} are very useful for
signal processing in communications, engineering, medicine, physics, etc.
Linear and bilinear transforms have been used. Among the linear transforms,
Fourier \cite{Fourier1888} and wavelets \cite{Combes90} {\cite{Daubechies90} 
\cite{Chui92} are the most popular. Among the bilinear ones, the
Wigner--Ville quasidistribution \cite{Wigner32} \cite{Ville48} provides
information in the joint time--frequency domain with good energy resolution.
A joint time--frequency description of signals is important, because in many
applications (biomedical, seismic, radar, etc.) the signals are of finite
(sometimes very short) duration. However, the oscillating cross-terms in the
Wigner--Ville quasidistribution make the interpretation of this transform a
difficult matter. Even if the average of the cross-terms is small, their
amplitude may be greater than the signal in time--frequency regions that
carry no physical information. To profit from the time--frequency energy
resolution of the bilinear transforms while controlling the cross-terms
problem, modifications to the Wigner--Ville transform have been proposed.
Transforms in the Cohen class \cite{Cohen1} \cite{Cohen2} make a
two-dimensional filtering of the Wigner--Ville quasidistribution and the
Gabor spectrogram \cite{Gabor} is a truncated version of this
quasidistribution. The difficulties with the physical interpretation of
quasidistributions arise from the fact that time and frequency correspond to
two noncommutative operators. Hence a joint probability density cannot be
defined.} Even in the case of positive quasiprobabilities like the
Husimi--Kano function \cite{Husimi} \cite{Kano}, an interpretation as a
joint probability distribution is also not possible because the two
arguments of the function are not simultaneously measurable random variables.

Recently, a new type of strictly positive bilinear transforms has been
proposed \cite{MendesPLA} \cite{MankoJPA}, called \textit{tomograms}, which
are a generalization of the Radon transform \cite{Radon} to noncommutative
pairs of operators. The Radon--Wigner transform \cite{Radon1} \cite{Radon2}
is a particular case of such noncommutative tomography technique. The
tomograms are strictly positive probability densities, provide a full
characterization of the signal and are robust in the presence of noise.

A unified framework to characterize linear transforms, quasidistributions
and tomograms was developed in Ref.\cite{MankoJPA}. This is briefly
summarized in Section 2. Then Sections 3,4,6 and 7 contains an explicit
construction of tomogram transforms for many pairs of noncommuting operators
in one and two dimensions. Some of these transforms have been used in the
past \cite{reflecto1} \cite{reflecto2}, others are completely new.

It is in the time-frequency plane that most signal processing experts have
developed their intuition, not in the eigenspaces associated to the new
tomograms. Therefore, to provide a qualitative intuition on the way the
tomograms explore the time-frequency plane, we have provided graphical
spectrograms of the eigenstates on which the signal is projected by the
tomograms.

In Section 5, an interpretation of the tomograms is given as operator
symbols of the set of projection operators in the space of signals. This
provides a very general framework to deal with all kinds of custom-designed
integral transforms both for deterministic and random signals. It also
provides an alternative framework for an algebraic formulation of signal
processing.

Finally, an illustration of how such transforms may be used to analyze
signals is contained in Section 8. A brief review of denoising, detection of
small signals and component separation, done in the past, is included as
well as an application of one of the new transforms.

\section{Linear transforms, quasi-distributions and tomograms}

Consider signals $f(t)$ as vectors $\mid f\rangle $ in a dense nuclear
subspace $\mathcal{N}$ of a Hilbert space $\mathcal{H}$ with dual space $%
\mathcal{N}^{*}$ (with the canonical identification $\mathcal{N\subset N}%
^{*} $) and a family of operators $\left\{ U(\alpha ):\alpha \in I, I \subset \mathbb{R}^n\right\} $
defined on $\mathcal{N}^{*}$ . In most cases of interest $U\left( \alpha
\right) $ generates a unitary group $U\left( \alpha \right) =e^{iB\left(
\alpha \right) }$. In this setting three types of integral transforms are
constructed.

Let $h\in \mathcal{N}^{*}$ be a reference vector and let $U$ be such that the linear span
of $\left\{ U(\alpha )h\in \mathcal{N}^{*}:\alpha \in I\right\} $ is dense
in $\mathcal{N}^{*}$ . In the set $\left\{ U(\alpha )h\right\} $, a complete
set of vectors can be chosen to serve as a basis.

\textbf{1 - Linear transforms} 
\begin{equation}
W_{f}^{(h)}(\alpha )=\langle U\left( \alpha \right) h\mid f\rangle
\label{2.1}
\end{equation}

\textbf{2 - Quasi-distributions} 
\begin{equation}
Q_{f}(\alpha )=\langle U\left( \alpha \right) f\mid f\rangle  \label{2.2}
\end{equation}

\textbf{3 - Tomograms}

If $U\left( \alpha \right) $ is a unitary operator there is a self-adjoint
operator $B\left( \alpha \right) $ such that $U\left( \alpha \right)
=e^{iB\left( \alpha \right) }$. The tomogram is 
\begin{equation}
M_{f}^{(B)}(X)=\langle f\mid \delta \left( B\left( \alpha \right) -X\right)
\mid f\rangle  \label{2.3}
\end{equation}

$X$ takes values on the spectrum of $B(\alpha )$. Considering a set of
generalized eigenstates (in $\mathcal{N}^{\ast }$) of $B(\alpha )$, one
obtains for the kernel 
\begin{equation*}
\langle Y\mid \delta \left( B(\alpha )-X\right) \mid Y^{\prime }\rangle
=\delta (Y^{\prime }-X)\,\delta (Y-Y^{\prime })=\langle Y\mid X\rangle
\langle X\mid Y^{\prime }\rangle
\end{equation*}%
Therefore, we may identify $\delta \left( B(\alpha )-X\right) $ with the
projector $\mid X\rangle \langle X\mid $ 
\begin{equation*}
\delta \left( B(\alpha )-X\right) =\mid X\rangle \langle X\mid =P_{X}
\end{equation*}%
From this, it follows 
\begin{equation}
M_{f}^{(B)}=\langle f\mid \delta \left( B(\alpha )-X\right) \mid f\rangle
=\langle f\mid X\rangle \langle X\mid f\rangle =|\langle X\mid f\rangle |^{2}
\label{2.4}
\end{equation}%
showing the positivity of the tomogram and its nature as the squared
amplitude of the projection on generalized eigenvectors of $B(\alpha )$.
Let, by a unitary transformation $S$, $B(\alpha )$ be transformed to%
\begin{equation*}
SB(\alpha )S^{\dagger }=B^{^{\prime }}(\alpha )
\end{equation*}%
If $\left\{ \mid Z\rangle \right\} $ is the set of (generalized)
eigenvectors of $B^{^{\prime }}(\alpha )$, $\left\{ S^{\dagger }\mid
Z\rangle \right\} $ is a set of eigenvectors for $B$. Therefore, 
\begin{equation*}
M_{f}^{(B)}(Z)=\langle f\mid \delta \left( B(\alpha )-Z\right) \mid f\rangle
=|\langle Z\mid S\mid f\rangle |^{2}=\langle f\mid S^{\dagger }\mid Z\rangle
\langle Z\mid S\mid f\rangle
\end{equation*}

For normalized $\mid f\rangle $, 
\begin{equation*}
\langle f\mid f\rangle =1
\end{equation*}
the tomogram is normalized 
\begin{equation}
\int M_{f}^{(B)}\left( X\right) \,dX=1  \label{2.5}
\end{equation}
It is a probability distribution for the random variable $X$ corresponding
to the observable defined by the operator $B\left( \alpha \right) $. The
tomogram is a homogeneous function 
\begin{equation}
M_{f}^{(B/p)}(X)=|p|M_{f}^{(B)}(pX)  \label{2.6}
\end{equation}

\textbf{Examples:}

If $U\left( \alpha \right) $ is unitary generated by $B_{F}\left( 
\overrightarrow{\alpha }\right) =\alpha _{1}t+i\alpha _{2}\frac{d}{dt}$ and $%
h$ is a (generalized) eigenvector of the time-translation operator the
linear transform $W_{f}^{(h)}(\alpha )$ is the Fourier transform. For the
same $B_{F}\left( \overrightarrow{\alpha }\right) $\textbf{,} the
quasi-distribution\textbf{\ }$Q_{f}(\alpha )$ is the ambiguity function.

The Wigner--Ville transform {\cite{Wigner32} \cite{Ville48}} is the
quasi-distribution $Q_{f}(\alpha )$ for the following $B-$operator 
\begin{equation}
B^{(WV)}(\alpha _{1},\alpha _{2})=-i2\alpha _{1}\frac{d}{dt}-2\alpha _{2}t+%
\frac{\pi \left( t^{2}-\frac{d^{2}}{dt^{2}}-1\right) }{2}\,  \label{2.7}
\end{equation}

The wavelet transform is $W_{f}^{(h)}(\alpha )$ for $B_{W}\left( 
\overrightarrow{\alpha }\right) =\alpha _{1}D+i\alpha _{2}\frac{d}{dt}$, $D$
being the dilation operator $D=-\frac{1}{2}\left( it\frac{d}{dt}+i\frac{d}{dt%
}t\right) $. The wavelets $h_{s,\,\tau }(t)$ are kernel functions generated
from a basic wavelet $h(\tau )$ by means of a translation and a rescaling $%
(-\infty <\tau <\infty ,$ $s>0)$: 
\begin{equation}
h_{s,\,\tau }(t)=\frac{1}{\sqrt{s}}\,h\left( \frac{t-\tau }{s}\right)
\label{2.8}
\end{equation}%
using the operator 
\begin{equation}
U^{(A)}(\tau ,s)=\exp (i\tau \hat{\omega})\exp (i\log \,sD),  \label{2.9}
\end{equation}%
\begin{equation}
h_{s,\tau }(t)=U^{(A)\dagger }(\tau ,s)h(t).  \label{2.10}
\end{equation}%
For normalized $h(t)$ the wavelets $h_{s,\,\tau }(t)$ satisfy the
normalization condition 
\begin{equation*}
\int |h_{s,\,\tau }(t)|^{2}\,dt=1.
\end{equation*}%
The basic wavelet (reference vector) may have different forms, for example, 
\begin{equation}
h(t)=\frac{1}{\sqrt{\pi }}\,e^{i\omega _{0}t}\,e^{-t^{2}/2},  \label{2.11a}
\end{equation}%
or 
\begin{equation}
h(t)=(1-t^{2})\,e^{-t^{2}/2}  \label{2.11b}
\end{equation}%
called the Mexican hat wavelet.

The Bertrand transform \cite{BerBerJMP} \cite{Baran} is $Q_{f}(\alpha )$ for 
$B_{W}\smallskip $.

Linear, bilinear and tomogram transforms are related to one another by

\begin{equation*}
M_{f}^{(B)}(X)=\frac{1}{2\pi }\int Q_{f}^{(kB)}(\alpha )\,e^{-ikX}\,dk
\end{equation*}

\begin{equation*}
Q_{f}^{(B)}(\alpha )=\int M_{f}^{(B/p)}(X)\,e^{ipX}\,dX
\end{equation*}

\begin{equation*}
Q_{f}^{(B)}(\alpha )=W_{f}^{(f)}(\alpha )
\end{equation*}

\begin{equation*}
W_{f}^{(h)}(\alpha )=\frac{1}{4}\int e^{iX}\left[ 
\begin{array}{c}
M_{f_{1}}^{(B)}(X)-iM_{f_{2}}^{(B)}(X) \\ 
-M_{f_{3}}^{(B)}(X)+iM_{f_{4}}^{(B)}(X)%
\end{array}%
\right] \,dX
\end{equation*}

with 
\begin{eqnarray*}
&\mid &f_{1}\rangle =\mid h\rangle +\mid f\rangle ;\qquad \mid f_{3}\rangle
=\mid h\rangle -\mid f\rangle \\
&\mid &f_{2}\rangle =\mid h\rangle +i\mid f\rangle ;\qquad \mid f_{4}\rangle
=\mid h\rangle -i\mid f\rangle
\end{eqnarray*}

\section{One-dimensional tomograms}

As shown in (\ref{2.4}) a tomogram corresponds to projections on the
eigenstates of the $B$ operators. These operators are linear combinations of
different (commuting or noncommuting) operators, 
\begin{equation*}
B=\mu O_{1}+\nu O_{2}
\end{equation*}%
Therefore the tomogram explores the signal along lines in the plane $\left(
O_{1},O_{2}\right) $. For example the tomogram%
\begin{equation}
M_{f}^{(S)}\left( X,\mu ,\nu \right) =\langle f\mid \delta \left( \mu t+\nu
\omega -X\right) \mid f\rangle  \label{3.1}
\end{equation}%
with $\omega =i\frac{d}{dt}$, is the expectation value of an operator
delta-function in the state $\mid f\rangle $, the support of the
delta-function being a line in the time--frequency plane 
\begin{equation}
X=\mu t+\nu \omega  \label{3.2}
\end{equation}%
Therefore, $M_{f}^{(S)}\left( X,\mu ,\nu \right) $ is the marginal
distribution of the variable $X$ along this line in the time--frequency
plane. The line is rotated and rescaled when one changes the parameters $\mu 
$ and $\nu $. In this way, the whole time--frequency plane is sampled and
the tomographic transform contains all the information on the signal.

It is clear that, instead of marginals collected along straight lines on the
time--frequency plane, one may use other curves to sample this space. It has
been shown in \cite{MankoJPA} that the tomograms associated to the affine
group, for example 
\begin{equation}
M_{f}^{(A_{t})}\left( X,\mu ,\nu \right) =\langle f\mid \delta \left( \mu
t+\nu \frac{t\omega +\omega t}{2}-X\right) \mid f\rangle  \label{3.3}
\end{equation}%
correspond to hyperbolas in the time-frequency plane. This point of view has
been further explored in \cite{Mankonew} defining tomograms in terms of
marginals over surfaces generated by deformations of families of hyperplanes
or quadrics. However not all tomograms may be defined as marginals on lines
in the time-frequency plane.

Here we construct the tomograms corresponding to a large set of operators
defined in terms of (one-dimensional) time. Of particular interest are the
tomograms associated to finite-dimensional Lie algebras.

\subsection{1D conformal group tomograms}

The generators of the one-dimensional conformal group are, 
\begin{equation}
\begin{array}{l}
\omega =i\frac{d}{dt} \\ 
D=i\left( t\frac{d}{dt}+\frac{1}{2}\right) \\ 
K=i\left( t^{2}\frac{d}{dt}+t\right)%
\end{array}
\label{3.4}
\end{equation}
One may construct t{omograms using the following operators:}

\textit{Time-frequency} 
\begin{equation}
B_{1}=\mu t+i\nu \frac{d}{dt}  \label{3.5}
\end{equation}

\textit{Time-scale} 
\begin{equation}
B_{2}=\mu t+i\nu \left( t\frac{d}{dt}+\frac{1}{2}\right)  \label{3.6}
\end{equation}

\textit{Frequency-scale} 
\begin{equation}
B_{3}=i\mu \frac{d}{dt}+i\nu \left( t\frac{d}{dt}+\frac{1}{2}\right)
\label{3.7}
\end{equation}

\textit{Time-conformal} 
\begin{equation}
B_{4}=\mu t+i\nu \left( t^{2}\frac{d}{dt}+t\right)  \label{3.8}
\end{equation}

The construction of the tomograms reduces to the calculation of the
generalized eigenvectors of each one of the $B_{i}$ operators

$B_{1}\psi _{1}\left( \mu ,\nu ,t,X\right) =X\psi _{1}\left( \mu ,\nu
,t,X\right) $%
\begin{equation}
\psi _{1}\left( \mu ,\nu ,t,X\right) =\exp i\left( \frac{\mu t^{2}}{2\nu }-%
\frac{tX}{\nu }\right)  \label{3.9}
\end{equation}
with normalization 
\begin{equation}
\int dt\psi _{1}^{*}\left( \mu ,\nu ,t,X\right) \psi _{1}\left( \mu ,\nu
,t,X^{\prime }\right) =2\pi \nu \delta \left( X-X^{\prime }\right)
\label{3.10}
\end{equation}

$B_{2}\psi _{2}\left( \mu ,\nu ,t,X\right) =X\psi _{2}\left( \mu ,\nu
,t,X\right) $%
\begin{equation}
\psi _{2}\left( \mu ,\nu ,t,X\right) =\frac{1}{\sqrt{\left| t\right| }}\exp
i\left( \frac{\mu t}{\nu }-\frac{X}{\nu }\log \left| t\right| \right)
\label{3.11}
\end{equation}
\begin{equation}
\int dt\psi _{2}^{*}\left( \mu ,\nu ,t,X\right) \psi _{2}\left( \mu ,\nu
,t,X^{\prime }\right) =4\pi \nu \delta \left( X-X^{\prime }\right)
\label{3.12}
\end{equation}

$B_{3}\psi _{3}\left( \mu ,\nu ,\omega ,X\right) =X\psi _{3}\left( \mu ,\nu
,\omega ,X\right) $%
\begin{equation}
\psi _{3}\left( \mu ,\nu ,t,X\right) =\exp \left( -i\right) \left( \frac{\mu 
}{\nu }\omega -\frac{X}{\nu }\log |\omega |\right)  \label{3.13}
\end{equation}
\begin{equation}
\int d\omega \psi _{1}^{*}\left( \mu ,\nu ,\omega ,X\right) \psi _{1}\left(
\mu ,\nu ,\omega ,X^{\prime }\right) =2\pi \nu \delta \left( X-X^{\prime
}\right)  \label{3.14}
\end{equation}

$B_{4}\psi _{4}\left( \mu ,\nu ,t,X\right) =X\psi _{4}\left( \mu ,\nu
,t,X\right) $%
\begin{equation}
\psi _{4}\left( \mu ,\nu ,t,X\right) =\frac{1}{\left| t\right| }\exp i\left( 
\frac{X}{\nu t}+\frac{\mu }{\nu }\log \left| t\right| \right)  \label{3.15}
\end{equation}
\begin{equation}
\int dt\psi _{4}^{*}\left( \mu ,\nu ,t,s\right) \psi _{4}\left( \mu ,\nu
,t,s^{\prime }\right) =2\pi \nu \delta \left( s-s^{\prime }\right)
\label{3.16}
\end{equation}

Then the tomograms are:

\textit{Time-frequency tomogram} 
\begin{equation}
M_{1}\left( \mu ,\nu ,X\right) =\frac{1}{2\,\pi |\nu |}\left| \int \exp %
\left[ \frac{i\mu t^{2}}{2\,\nu }-\frac{itX}{\nu }\right] f(t)\,dt\right|
^{2}  \label{3.17}
\end{equation}

\textit{Time-scale tomogram} 
\begin{equation}
M_{2}(\mu ,\nu ,X)=\frac{1}{2\pi |\nu |}\left| \int dt\,\frac{f(t)}{\sqrt{|t|%
}}e^{\left[ i\left( \frac{\mu }{\nu }t-\frac{X}{\nu }\log |t|\right) \right]
}\right| ^{2}  \label{3.18}
\end{equation}

\textit{Frequency-scale tomogram} 
\begin{equation}
M_{3}(\mu ,\nu ,X)=\frac{1}{2\pi |\nu |}\left| \int d\omega \,\frac{f(\omega
)}{\sqrt{|\omega |}}e^{\left[ -i\left( \frac{\mu }{\nu }\omega -\frac{X}{\nu 
}\log |\omega |\right) \right] }\right| ^{2}  \label{3.19}
\end{equation}
$f(\omega )$ being the Fourier transform of $f(t)$

\begin{figure} [htb]
\centering
\includegraphics[width=12cm]{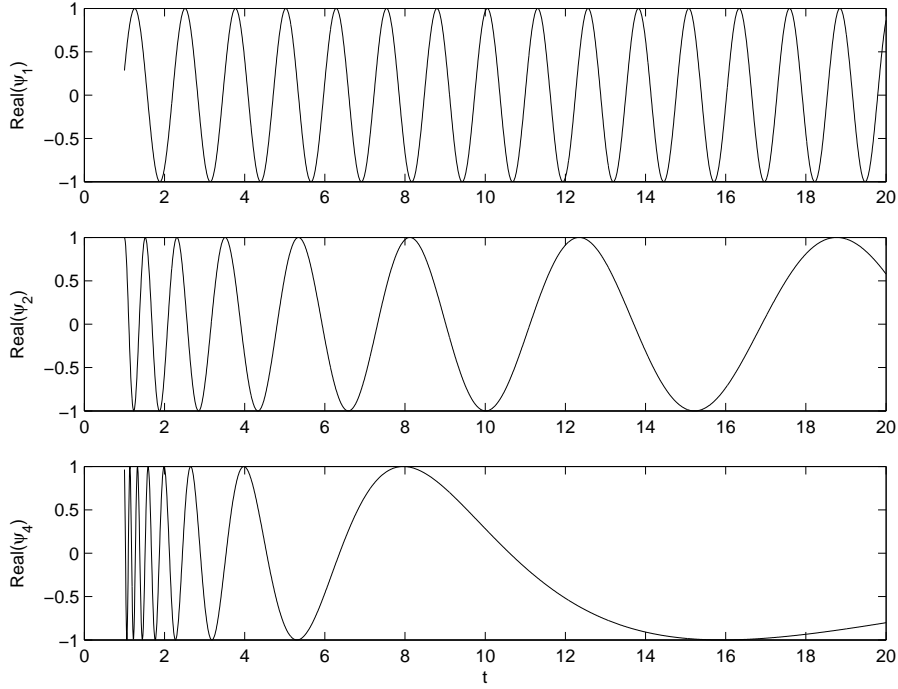} 
\caption{Typical behaviour of the real
part of the functions $\protect\psi _{1}$,$\protect\psi _{2}$ and $\protect%
\psi _{4}$ at $\protect\mu =0$}
\label{realpart}
\end{figure}

\textit{Time-conformal tomogram} 
\begin{equation}
M_{4}(\mu ,\nu ,X)=\frac{1}{2\pi |\nu |}\left\vert \int dt\,\frac{f(t)}{|t|}%
e^{\left[ i\left( \frac{X}{\nu t}+\frac{\mu }{\nu }\log |t|\right) \right]
}\right\vert ^{2}  \label{3.20}
\end{equation}%
The tomograms $M_{1},M_{2}$ and $M_{4}$ interpolate between the (squared)
time signal and its projection on the $\psi _{i}\left( \mu ,\nu ,t,X\right) $
functions for $\mu =0$. Fig.\ref{realpart} shows the typical behavior of the real part of
these functions.

\begin{figure}[tbp]
\centering
\includegraphics[width=12cm]{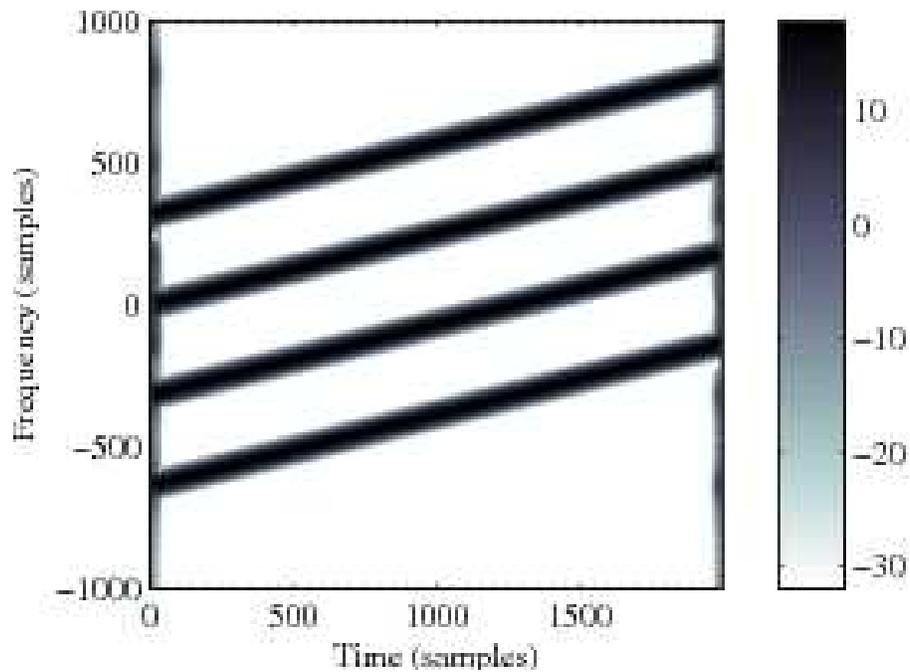} 
\caption{Modulus of the Short-time Fourier transform of 4 vectors of the
time-frequency tomogram for some fixed $\protect\theta$, $\protect\mu=\cos%
\protect\theta$, $\protect\nu=\sin\protect\theta$. A vector is a linear
chirp, hence a line in the time-frequency plane. Moreover, each vector is a
frequency-translated version of the one which starts at the origin. Since it
forms an orthogonal basis, the sum of all the vectors cover the entire
time-frequency plane. The parameter $\protect\theta$ allows to change the
slope of the line in the time-frequency plane.}
\label{timefreq}
\end{figure}

\begin{figure}[tbp]
\centering
\begin{minipage}[c]{.46\linewidth}
      \includegraphics[width=7cm]{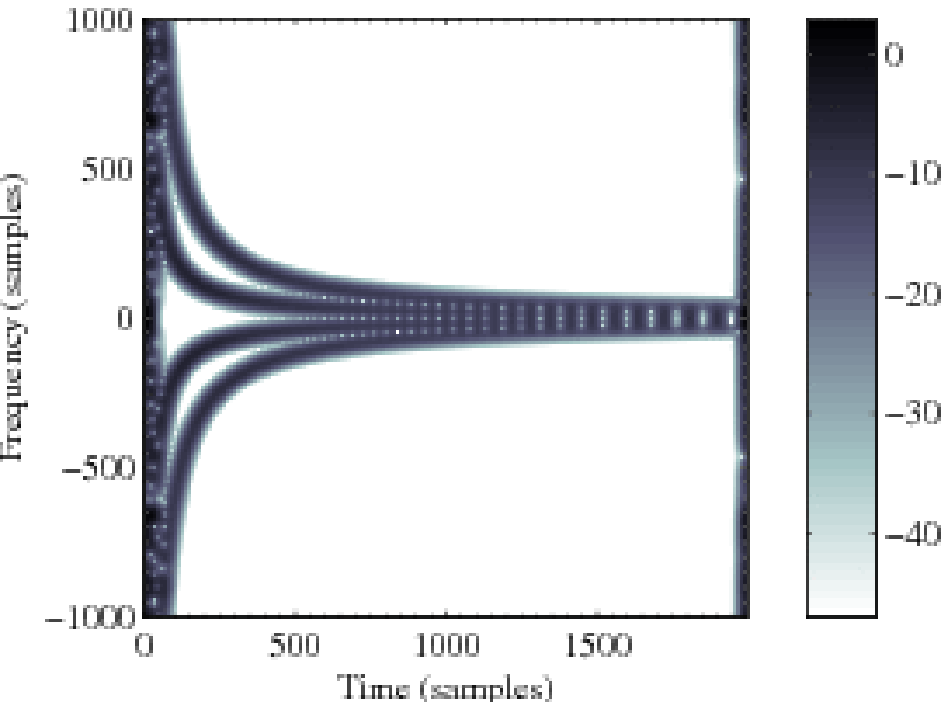}
   \end{minipage} \hfill  
\begin{minipage}[c]{.46\linewidth}
      \includegraphics[width=7cm]{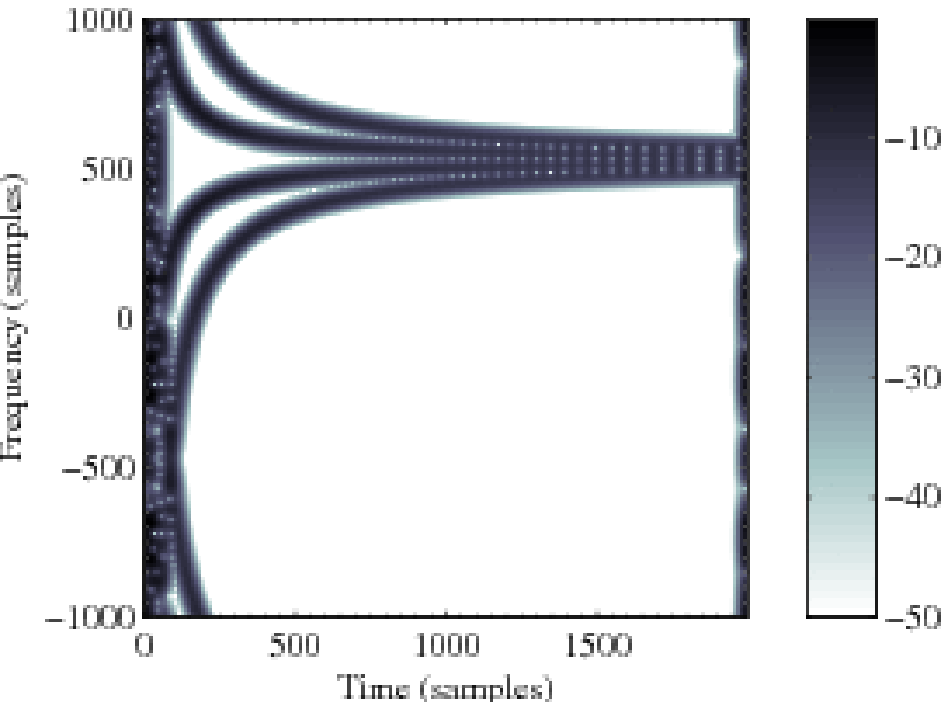}
   \end{minipage}
\caption{Modulus of the Short-time Fourier transform of 4 vectors of the
time-scale tomogram for $\protect\mu=0$, $\protect\nu=1$ (left) and $\protect%
\mu=\protect\sqrt(2)/2$, $\protect\nu=\protect\sqrt(2)/2$ (right). Each
vector is an hyperbolic chirp. Two of them correspond to positive $X$ and
two of them to negative $X$. Due to the sampling used in the numerical
computation, some aliasing phenomenum occurs at times close to zero. There
is a axis of symetry: the line of zero frequency on the left graph. This
axis is shifted in frequency when $\protect\mu$ and $\protect\nu$ are
changed.}
\label{timescale}
\end{figure}

\begin{figure}[tbp]
\centering
\includegraphics[width=12cm]{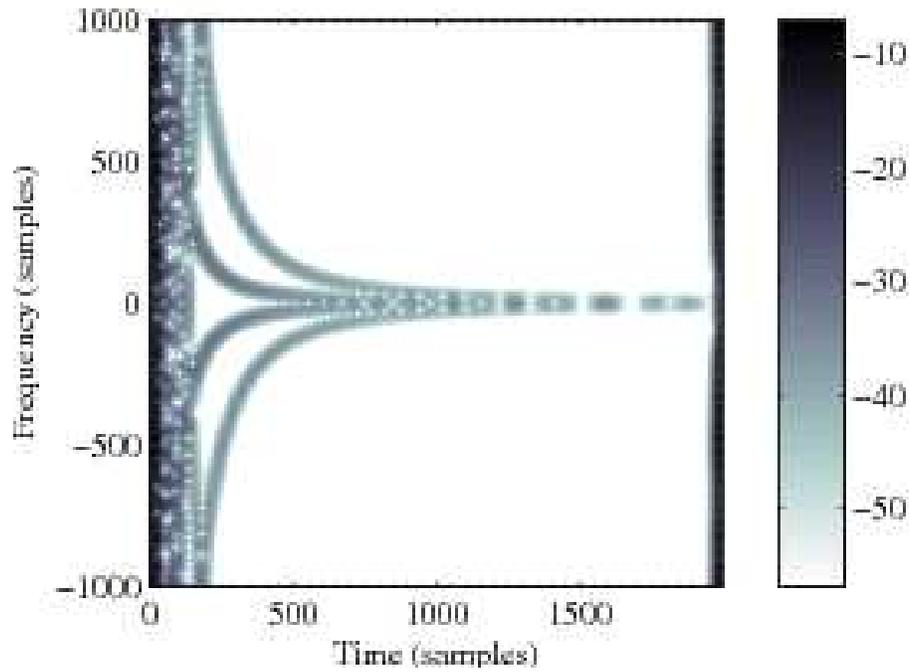} 
\caption{Modulus of the Short-time Fourier transform of 4 vectors of the
time-conformal tomogram for $\protect\mu=0$, $\protect\nu=1$. Due to the
sampling used in the numerical computation, some aliasing phenomenum occurs
at times close to zero. Some interferences between the vectors occur for
large time. Two vectors correspond to positive $X$ and two to negative $X$.}
\label{conformal}
\end{figure}

Figs.\ref{timefreq},\ref{timescale} and \ref{conformal} illustrate how the tomograms $M_{1},M_{2}$ and $M_{4}$
explore the time-frequency space by plotting the spectrograms of typical
vectors $\psi _{1},\psi _{2}$ and $\psi _{4}$. 

In a similar way, tomograms may be constructed for any operator of the
general type 
\begin{equation*}
B_{4}=\mu t+i\nu \left( g\left( t\right) \frac{d}{dt}+\frac{1}{2}\frac{%
dg\left( t\right) }{dt}\right)
\end{equation*}%
the generalized eigenvectors being 
\begin{equation*}
\psi _{g}\left( \mu ,\nu ,t,X\right) =\left\vert g\left( t\right)
\right\vert ^{-1/2}\exp i\left( -\frac{X}{\nu }\int^{t}\frac{ds}{g\left(
s\right) }+\frac{\mu }{\nu }\int^{t}\frac{sds}{g\left( s\right) }\right)
\end{equation*}

\subsection{Another finite-dimensional algebra}

Another finite-dimensional Lie algebra which may be used to construct
tomograms, exploring other features of the signals, is generated by $\mathbf{%
1}$, $t$ and

\begin{equation*}
\begin{array}{l}
\omega =i\frac{d}{dt} \\ 
D=i\left( t\frac{d}{dt}+\frac{1}{2}\right) \\ 
F=-\frac{1}{2}\left( \frac{d^{2}}{dt^{2}}-t^{2}+1\right) \\ 
\sigma =\frac{1}{2}\left( \frac{d^{2}}{dt^{2}}+t^{2}+1\right)%
\end{array}%
\end{equation*}

Of special interest are the tomograms related to the operators 
\begin{equation*}
B_{F}=\mu t+\nu F
\end{equation*}
and 
\begin{equation*}
B_{\sigma }=\mu t+\nu \sigma
\end{equation*}

As before, the construction of the tomograms relies on finding a complete
set of generalized eigenvectors for the operators $B_{F}$ and $B_{\sigma }$.
With $y=t+\frac{\mu }{\nu }$ one defines creation and annihilation operators 
\begin{eqnarray*}
a &=&\frac{1}{\sqrt{2}}\left( y+\frac{d}{dy}\right) \\
a^{\dagger } &=&\frac{1}{\sqrt{2}}\left( y-\frac{d}{dy}\right)
\end{eqnarray*}%
obtaining 
\begin{equation*}
B_{F}=\nu \left( a^{\dagger }a-\frac{\mu ^{2}}{2\nu ^{2}}\right)
\end{equation*}%
\begin{equation*}
B_{\sigma }=\nu \left( aa-\frac{\mu ^{2}}{2\nu ^{2}}\right)
\end{equation*}%
Therefore for $B_{F}$ one has an orthonormalized complete set of
eigenvectors 
\begin{equation*}
\psi _{n}^{(F)}\left( t\right) =u_{n}\left( t+\frac{\mu }{\nu }\right)
\end{equation*}%
with a discrete set of eigenvalues $X_{n}=\nu \left( n+\frac{1}{2}\right) -%
\frac{\mu ^{2}}{2\nu }$%
\begin{equation*}
B_{F}\psi _{n}^{(F)}\left( t\right) =X_{n}\psi _{n}^{(F)}\left( t\right)
\end{equation*}%
the function $u_{n}$ being 
\begin{equation*}
u_{n}\left( y\right) =\left( \pi ^{1/2}2^{n}n!\right) ^{-1/2}\left( y-\frac{d%
}{dy}\right) ^{n}e^{-\frac{y^{2}}{2}}
\end{equation*}%
The tomogram $M_{f}^{(F)}\left( \mu ,\nu ,X_{n}\right) $ is 
\begin{equation*}
M_{f}^{(F)}\left( \mu ,\nu ,X_{n}\right) =\left\vert \int \psi _{n}^{(F)\ast
}\left( t\right) f\left( t\right) dt\right\vert ^{2}
\end{equation*}

For $B_{\sigma }$ one uses a basis of coherent states 
\begin{eqnarray*}
\phi _{\lambda }\left( y\right) &=&e^{\lambda a^{\dagger }-\lambda ^{\ast
}a}u_{0}\left( y\right) \\
&=&e^{\frac{\left\vert \lambda \right\vert ^{2}}{2}}\sum_{n=0}\frac{\lambda
^{n}}{\sqrt{n!}}u_{n}\left( y\right)
\end{eqnarray*}%
with decomposition of identity 
\begin{equation*}
\frac{1}{\pi }\int \phi _{\lambda }\left( y\right) \phi _{\lambda }^{\ast
}\left( y\right) d^{2}\lambda =1
\end{equation*}%
Then, a set of generalized eigenstates of $B_{\sigma }$ is 
\begin{equation*}
\psi _{\lambda }^{(\sigma )}\left( \mu ,\nu ,t\right) =\phi _{\lambda
}\left( t+\frac{\mu }{\nu }\right)
\end{equation*}%
with eigenvalues 
\begin{equation*}
B_{\sigma }\psi _{\lambda }^{(\sigma )}\left( \mu ,\nu ,t\right) =X_{\lambda
}\psi _{\lambda }^{(\sigma )}\left( \mu ,\nu ,t\right)
\end{equation*}%
\begin{equation*}
X_{\lambda }=\nu \left( \lambda ^{2}-\frac{\mu ^{2}}{2\nu ^{2}}\right)
\end{equation*}%
the tomogram being 
\begin{equation*}
M_{f}^{\left( \sigma \right) }\left( \mu ,\nu ,X_{\lambda }\right)
=\left\vert \int \psi _{\lambda }^{(\sigma )\ast }\left( \mu ,\nu ,t\right)
f\left( t\right) dt\right\vert ^{2}
\end{equation*}%
This tomogram is closely related to the Sudarshan-Glauber P-representation 
\cite{Sudarshan} \cite{Glauber63}.

\section{Multidimensional tomograms}

Several types of multidimensional tomograms may be obtained from
generalizations of the one-dimensional ones. Consider a signal $%
f(t_{1},t_{2})$. The tomogram will depend on a vector variable $\vec{X}%
=\left( X_{1},X_{2}\right) $ and four real parameters $\mu _{1}$, $\mu _{2}$,%
$\nu _{1}$, and $\nu _{2}$. For example, the two-dimensional time-frequency
tomogram will be 
\begin{equation}
M(\vec{X},\vec{\mu},\vec{\nu})=\frac{1}{2\pi |\nu _{1}|}\,\frac{1}{2\pi |\nu
_{2}|}\left\vert \int f(t_{1},t_{2})\exp \left( \frac{i\mu _{1}}{2\nu _{1}}%
\,t_{1}^{2}-\frac{iX_{1}}{\nu _{1}}\,t_{1}+\frac{i\mu _{2}}{2\nu _{2}}%
\,t_{2}^{2}-\frac{iX_{2}}{\nu _{2}}\,t_{2}\right)
\,dt_{1}\,dt_{2}\right\vert ^{2}  \label{Mw}
\end{equation}%
From this one may also construct a \textit{center of mass tomogram} 
\begin{eqnarray*}
&&M_{\mathrm{cm}}(Y,\vec{\mu},\vec{\nu})=\int M(\vec{X},\vec{\mu},\vec{\nu}%
)\,\delta (Y-X_{1}-X_{2})\,dX_{1}\,dX_{2}=\int \delta (Y-X_{1}-X_{2})\,\frac{%
1}{2\pi |\nu _{1}|}\,\frac{1}{2\pi |\nu _{2}|} \\
&&\times \left\vert \int f(t_{1},t_{2})dt_{1}\,dt_{2}\,\exp \left( \frac{%
i\mu _{1}}{2\nu _{1}}\,t_{1}^{2}-\frac{iz_{1}X_{1}}{\nu _{1}}+\frac{i\mu _{2}%
}{2\nu _{2}}\,t_{2}^{2}-\frac{iz_{2}X_{2}}{\nu _{2}}\right) \right\vert
^{2}\,dX_{1}\,dX_{2}
\end{eqnarray*}%
the center of mass tomogram being normalized 
\begin{equation*}
\int M_{\mathrm{cm}}(X,\vec{\mu},\vec{\nu})\,dX=1
\end{equation*}%
and a homogeneous function 
\begin{equation*}
M_{\mathrm{cm}}(\lambda X,\lambda \vec{\mu},\lambda \vec{\nu})=\frac{1}{%
|\lambda |}\,M_{\mathrm{cm}}(X,\vec{\mu},\vec{\nu}).
\end{equation*}%
The generalization to $N$ channels is straightforward.

As in the one-dimensional case, useful tomograms may be constructed from the
operators of Lie algebras. For example, given the generators of the
conformal algebra in $\mathbb{R}^{d}$, $d\geq 2$, 
\begin{equation*}
\begin{array}{l}
\omega _{k}=i\frac{\partial }{\partial t_{k}} \\ 
D=i\left( t\bullet \nabla +\frac{d}{2}\right) \\ 
R_{j,k}=i\left( t_{j}\frac{\partial }{\partial t_{k}}-t_{k}\frac{\partial }{%
\partial t_{j}}\right) \\ 
K_{j}=i\left( t_{j}^{2}\frac{\partial }{\partial t_{j}}+t_{j}\right)%
\end{array}%
\end{equation*}%
Let, in two dimensions, $t_{1}=t$ and $t_{2}=x$. The tomograms corresponding
to the operators 
\begin{eqnarray*}
B_{\omega } &=&\mu _{1}t+\mu _{2}x+\nu _{1}\omega _{1}+\nu _{2}\omega _{2} \\
B_{D} &=&\mu _{1}t+\mu _{2}x+\nu D \\
B_{\omega } &=&\mu _{1}t+\mu _{2}x+\nu _{1}K_{1}+\nu _{2}K_{2}
\end{eqnarray*}%
are, as in (\ref{Mw}), straightforward generalizations of the corresponding
one-dimensional ones. For the operator 
\begin{equation*}
B_{R}=\mu _{1}t+\mu _{2}x+\nu R_{1,2}
\end{equation*}%
the eigenstates are 
\begin{equation*}
\psi ^{(R)}\left( \overset{\rightarrow }{\mu },\nu ,x,t,X\right) =\exp \frac{%
i}{\nu }\left( \mu _{1}x-\mu _{2}t+X\tan ^{-1}\frac{t}{x}\right)
\end{equation*}%
and the tomogram 
\begin{equation*}
M_{f}\left( \overset{\rightarrow }{\mu },\nu ,X\right) =\left\vert \int \psi
^{(R)\ast }\left( \overset{\rightarrow }{\mu },\nu ,x,t,X\right) f\left(
x,t\right) dxdt\right\vert ^{2}
\end{equation*}

\section{The tomograms as operator symbols}

Tomograms may be described not only as amplitudes of projections on a
complete basis of eigenvectors of a family of operators, but also as
operator symbols. That is, as a map of operators to a space of functions
where the operators non-commutativity is replaced by a modification of the
usual product to a star-product.

Let $\hat{A}$ be an operator in Hilbert space $\mathcal{H}$ and $\hat{U}(%
\vec{x})$, $\hat{D}(\vec{x})$ two families of operators called \textit{%
dequantizers} and \textit{quantizers}, respectively, such that 
\begin{equation}
\text{Tr}\left\{ \,\hat{U}(\vec{x})\hat{D}(\vec{x}^{\prime })\right\}
=\delta (\vec{x}-\vec{x}^{\prime })  \label{5.2}
\end{equation}%
The labels $\vec{x}$ (with components $x_{1},x_{2},\ldots x_{n}$) are
coordinates in a linear space $V$ where the functions (operator symbols) are
defined. Some of the coordinates may take discrete values, then the delta
function in (\ref{5.2}) should be understood as a Kronecker delta. Provided
the property (\ref{5.2}) is satisfied, one defines the \textit{symbol of the
operator} $\hat{A}$ by the formula 
\begin{equation}
f_{A}(\vec{x})=\text{Tr}\left\{ \hat{U}(\vec{x})\hat{A}\right\} ,
\label{5.3}
\end{equation}%
assuming the trace to exist. In view of (\ref{5.2}), one has the
reconstruction formula 
\begin{equation}
\hat{A}=\int f_{A}(x)\hat{D}(\vec{x})\,d\vec{x}  \label{5.4}
\end{equation}%
The role of quantizers and dequantizers may be exchanged. Then 
\begin{equation}
f_{A}^{d}(\vec{x})=\text{Tr}\left\{ \hat{D}(\vec{x})\,\hat{A}\right\}
\label{5.6}
\end{equation}%
is called the dual symbol of $f_{A}(\vec{x})$ and the reconstruction formula
is 
\begin{equation}
\hat{A}=\int f_{A}^{d}(x)\hat{U}(\vec{x})\,d\vec{x}  \label{5.7}
\end{equation}%
Symbols of operators can be multiplied using the star-product kernel as
follows 
\begin{equation}
f_{A}(\vec{x})\star f_{B}(\vec{x})=\int f_{A}(\vec{y})f_{B}(\vec{z})K(\vec{y}%
,\vec{z},\vec{x})\,d\vec{y}\,d\vec{z}  \label{5.9}
\end{equation}%
the kernel being%
\begin{equation}
K(\vec{y},\vec{z},\vec{x})=\text{Tr}\left\{ \hat{D}(\vec{y})\hat{D}(\vec{z})%
\hat{U}(\vec{x})\right\}  \label{5.10}
\end{equation}%
The star-product is associative, 
\begin{equation}
\left( f_{A}(\vec{x})\star f_{B}(\vec{x})\right) \star f_{C}(\vec{x})=f_{A}(%
\vec{x})\star \left( f_{B}(\vec{x})\star f_{C}(\vec{x})\right)  \label{5.11}
\end{equation}%
this property corresponding to the associativity of the product of operators
in Hilbert space.

With the dual symbols the trace of an operator may be written in integral
form 
\begin{equation}
\text{Tr}\left\{ \,\hat{A}\hat{B}\right\} =\int f_{A}^{d}(\vec{x})f_{B}(\vec{%
x})\,d\vec{x}=\int f_{B}^{d}(\vec{x})f_{A}(\vec{x})\,d\vec{x}.  \label{5.13}
\end{equation}

For two different symbols $f_{A}(\vec{x})$ and $f_{A}(\vec{y})$
corresponding, respectively, to the pairs ($\hat{U}(\vec{x})$,$\hat{D}(\vec{x%
})$) and ($\hat{U}_{1}(\vec{y})$,$\hat{D}_{1}(\vec{y})$), one has the
relation 
\begin{equation}
f_{A}(\vec{x})=\int f_{A}(\vec{y})K(\vec{x},\vec{y})\,d\vec{y},  \label{5.14}
\end{equation}%
with intertwining kernel 
\begin{equation}
K(\vec{x},\vec{y})=\text{Tr}\left\{ \hat{D}_{1}(\vec{y})\hat{U}(\vec{x}%
)\right\}  \label{5.15}
\end{equation}

Let now each signal $f\left( t\right) $ be identified with the projection
operator $\Pi _{f}$ on the function $f\left( t\right) $, denoted by%
\begin{equation}
\Pi _{f}=\left\vert f\right\rangle \left\langle f\right\vert  \label{5.15a}
\end{equation}%
Then the tomograms and also other transforms are symbols of the projection
operators for several choices of quantizers and dequantizers.

Some examples:

\# The \textit{Wigner-Ville function}: is the symbol of $\mid f\rangle
\langle f\mid $ corresponding to the dequantizer 
\begin{equation}
\hat{U}(\vec{x})=2\hat{\mathcal{D}}(2\alpha )\hat{P},\qquad \alpha =\frac{%
t+i\omega }{\sqrt{2}}\,  \label{5.16}
\end{equation}%
where $\hat{P}$ is the inversion operator 
\begin{equation}
\hat{P}f(t)=f(-t)  \label{5.18}
\end{equation}%
and $\hat{\mathcal{D}}(\gamma )$ is a \textquotedblleft
displacement\textquotedblright\ operator 
\begin{equation}
\hat{\mathcal{D}}(\gamma )=\exp \left[ \frac{1}{\sqrt{2}}\gamma \left( t-%
\frac{\partial }{\partial t}\right) -\frac{1}{\sqrt{2}}\gamma ^{\ast }\left(
t+\frac{\partial }{\partial t}\right) \right]  \label{5.19}
\end{equation}%
The quantizer operator is 
\begin{equation}
\hat{D}(\vec{x}):=\hat{D}(t,\omega )=\frac{1}{2\pi }\hat{U}(t,\omega ),
\label{5.20}
\end{equation}%
$t$ and $\omega $ being time and frequency.

The Wigner--Ville function is 
\begin{equation}
W(t,\omega )=2\text{Tr}\left\{ \mid f\rangle \langle f\mid \hat{D}(2\alpha )%
\hat{D}\right\}  \label{5.21}
\end{equation}%
or, in integral form 
\begin{equation}
W(t,\omega )=2\int f^{\ast }(t)\hat{\mathcal{D}}(2\alpha )f(-t)\,dt
\label{5.22}
\end{equation}

\# The \textit{symplectic tomogram} or time-frequency tomogram of $\mid
f\rangle \langle f\mid $ corresponds to the dequantizer 
\begin{equation}
\hat{U}(\vec{x}):=\hat{U}(X,\mu ,\nu )=\delta \left( X\hat{1}-\mu \hat{t}%
-\nu \hat{\omega}\right) ,  \label{5.23}
\end{equation}%
with 
\begin{equation}
\hat{t}f(t)=tf(t),\qquad \hat{\omega}f(t)=-i\frac{\partial }{\partial t}%
\,f(t)  \label{5.24}
\end{equation}%
and $X,\mu ,\nu \in R$. The quantizer of the symplectic tomogram is 
\begin{equation}
\hat{D}(\vec{x}):=\hat{D}(X,\mu ,\nu )=\frac{1}{2\pi }\,\exp \left[ i\left( X%
\hat{1}-\mu \hat{t}-\nu \hat{\omega}\right) \right]  \label{5.25}
\end{equation}

\# The \textit{optical tomogram} is the same as above for the case 
\begin{equation}
\mu =\cos \theta ,\qquad \nu =\sin \theta .  \label{5.26}
\end{equation}%
Thus the optical tomogram is 
\begin{eqnarray}
M(X,\theta ) &=&\text{Tr}\left\{ \mid f\rangle \langle f\mid \delta \left( X%
\hat{1}-\mu \hat{t}-\nu \hat{\omega}\right) \right\}  \notag \\
&=&\frac{1}{2\pi }\int f^{\ast }(t)e^{ikX}\exp \left[ ik\left( X-t\cos
\theta +i\,\frac{\partial }{\partial t}\sin \theta \right) \right]
f(t)\,dt\,dk  \notag \\
&=&\frac{1}{2\pi |\sin \theta |}\left\vert \int f(t)\exp \left[ i\left( 
\frac{\cot \theta }{2}t^{2}-\frac{Xt}{\sin \theta }\right) \right]
dt\right\vert ^{2}.  \label{5.28}
\end{eqnarray}

One important feature of the formulation of tomograms as operator symbols is
that one may work with deterministic signals $f\left( t\right) $ as easily
as with probabilistic ones. In this latter case the projector in (\ref{5.15a}%
) would be replaced by%
\begin{equation}
\Pi _{p}=\int p_{\mu }\left\vert f_{\mu }\right\rangle \left\langle f_{\mu
}\right\vert d\mu  \label{5.29}
\end{equation}%
with $\int p_{\mu }d\mu =1$, the tomogram being the symbol of this new
operator.

This also provides a framework for an algebraic formulation of signal
processing, perhaps more general than the one described in \cite{Moura1} 
\cite{Moura2}. There, a signal model is a triple $\left( \mathcal{A},%
\mathcal{M},\Phi \right) $ $\mathcal{A}\ $being an algebra of linear
filters, $\mathcal{M}$ a $\mathcal{A}$-module and $\Phi $ a map from the
vector space of signals to the module. With the operator symbol
interpretation both (deterministic or random) signals and (linear or
nonlinear) transformations on signals are operators. By the application of
the dequantizer (Eq. \ref{5.3}) they are mapped onto functions, the filter
operations becoming star-products.

\section{Rotated-time tomography}

Now we consider a version of tomography where a discrete random variable is
used as an argument of the probability distribution function. We call this
tomography \textit{rotated time tomography}. It is a variant of the
spin-tomographic approach for the description of discrete spin states in
quantum mechanics. For a finite duration signal $f(t)$, with $0\leq t\leq T$%
,we consider discrete values of time $f(t_{m})\equiv f_{m}$, where with the
labeling $m=-j,-j+1,-j+2,\ldots ,0,1,\ldots ,j-1,j$ they are like the
components of a spinor $\mid f\rangle $. This means that we split the
interval $[0,T]$ onto $N$ parts at time values $t_{-j},t_{-j+1},\ldots
,t_{j} $ and replace the signal $f(t)$, a function of continuous time, by a
discrete set of values organized as a spinor. By dividing by a factor we
normalize the spinor, i.e., 
\begin{equation}
\langle f\mid f\rangle =\sum_{m=-j}^{j}|f_{m}|^{2}=1  \label{S1}
\end{equation}%
Without loss of generality, we consider the "spin" values to be integers,
i.e., $j=0,1,2,\ldots $ and use an odd number $N=2j+1$ of values.

In this setting, $\mid f\rangle $ being a column vector, we construct the $N$%
$\times $$N$ matrix 
\begin{equation}
\rho =\mid f\rangle \langle f\mid  \label{S2}
\end{equation}%
with matrix elements 
\begin{equation}
\rho _{mm^{\prime }}=f_{m}f_{m^{\prime }}^{\ast }.  \label{S3}
\end{equation}%
The tomogram is defined as the probability-distribution function 
\begin{equation}
\mathcal{M}(m,u)=|\langle m\mid u\mid f\rangle |^{2},\qquad m=-j,\ldots
,j-1,j  \label{S4}
\end{equation}%
where $u$ is the unitary $N$$\times $$N$ matrix 
\begin{equation}
uu^{\dagger }=1_{N}  \label{S5}
\end{equation}%
For this matrix we use an unitary irreducible representation of the rotation
group (or $SU(2)$) with matrix elements 
\begin{eqnarray}
u_{mm^{\prime }}(\theta ) &=&\frac{(-1)^{j-m^{\prime }}}{(m+m^{\prime })!}%
\left[ \frac{(j+m)!(j+m^{\prime })!}{(j-m)!(j-m^{\prime })!}\right]
^{1/2}\left( \sin \frac{\theta }{2}\right) ^{m-m^{\prime }}\left( \cos \frac{%
\theta }{2}\right) ^{m+m^{\prime }}  \notag \\
&&\times \mathcal{F}_{j-m}\left( 2m+1,m+m^{\prime 2}\frac{\theta }{2}\right)
\label{S61}
\end{eqnarray}%
$\mathcal{F}_{j-m}$ being a function with Jacobi polynomial structure
expressed in terms of hypergeometric function as 
\begin{eqnarray}
\mathcal{F}_{n}(a,b,t) &=&F(-n,a+n,b;t)=\frac{(b-1)!}{(b+n-1)!}%
\,t^{1-b}(1-t)^{b-a}\left( \frac{d}{dt}\right) ^{n}\left[
t^{b+n-1}(1-t)^{a-b+1}\right]  \notag \\
&&  \label{S71}
\end{eqnarray}%
The dequantizer in the rotated-time tomography is 
\begin{equation}
\hat{U}(\vec{x})\equiv U(m,\vec{n})=\delta (m1-u^{\dagger }J_{z}u)=\delta
\left( m1-\vec{n}\vec{J}\right)  \label{S8}
\end{equation}%
where $J_{z}$ is the matrix with diagonal matrix elements 
\begin{equation}
(J_{z})_{mm^{\prime }}=m\delta _{mm^{\prime }}  \label{S9}
\end{equation}%
The vector $\vec{n}=(\sin \theta \cos \varphi ,\sin \theta \sin \varphi
,\cos \theta )$ determines a direction in 3D space. The matrix (\ref{S61})
was written for $\varphi =0$ but, if this angle is nonzero, the matrix
element has to be multiplied by the phase factor $e^{im\varphi }$.

The quantizer can take several forms:

In integral form, it reads 
\begin{equation}
\hat{D}(m,\vec{n})=\frac{2j+1}{\pi }\int_{0}^{2\pi }\sin ^{2}\frac{\gamma }{2%
}\exp (-i\vec{J}\vec{n})\gamma \,d\gamma (\cdots )  \label{S10}
\end{equation}%
The tomogram $\mathcal{M}(m,u)$ is a nonnegative normalized probability
distribution depending on the direction $\vec{n}$, i.e., $\mathcal{M}%
(m,u)\geq 0$ and 
\begin{equation}
\sum_{m=-j}^{j}\mathcal{M}(m,u)=1  \label{S11}
\end{equation}%
To compute the tomogram for a given direction with angles $\varphi =0$ and $%
\theta $, one has to estimate 
\begin{equation}
\mathcal{M}(m,\theta )=\sum_{m^{\prime \prime },m^{\prime
}=-j}^{j}u_{mm^{\prime }}^{\ast }(\theta )f_{m}f_{m^{\prime \prime }}^{\ast
}u_{m^{\prime \prime }m}(\theta )  \label{S12}
\end{equation}%
where the matrix $u_{m^{\prime \prime }m}(\theta )$ is given by (\ref{S61}).
The following form for the matrix $u_{m^{\prime }m}(\theta )$ is more
convenient for numerical calculations: 
\begin{equation}
u_{m^{\prime }m}(\theta )=\left[ \frac{(j+m^{\prime })!(j-m^{\prime })!}{%
(j+m)!(j-m)!}\right] ^{1/2}\left( \cos \frac{\theta }{2}\right) ^{m^{\prime
}+m}\left( \sin \frac{\theta }{2}\right) ^{m^{\prime }-m}P_{j-m^{\prime
}}^{m^{\prime }-m,m^{\prime }+m}(\cos \theta )  \label{LL}
\end{equation}%
where $P_{n}^{a,b}$ are Jacobi polynomials.

In principle, one could use not only the unitary matrix in (\ref{S61}) but
arbitrary unitary matrices. They contain a larger number of parameters
(equal to $N^{2}-1)$ and can provide additional information on the signal
structure.

\begin{figure}
\centering
\includegraphics[width=12cm]{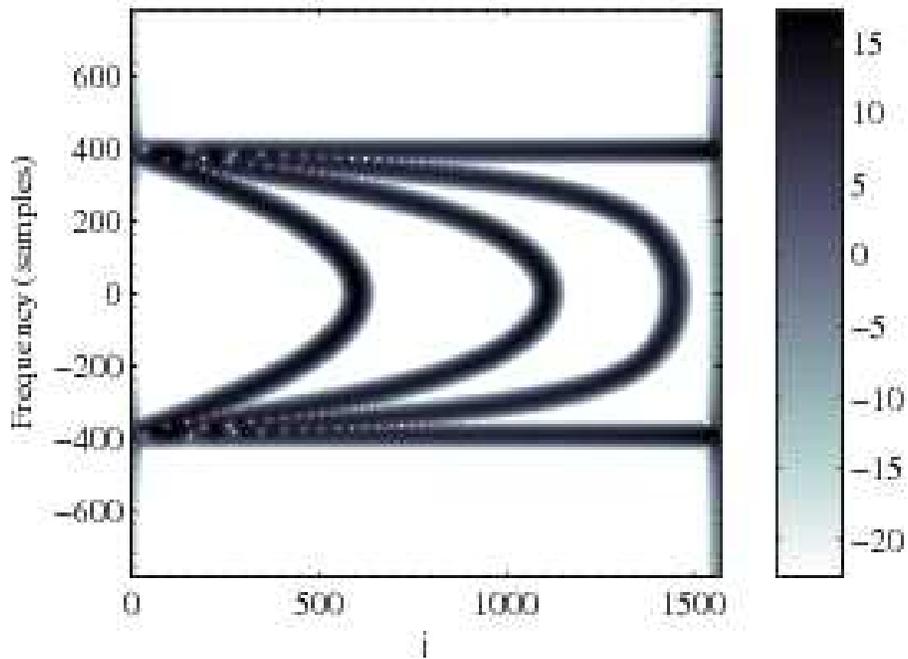}
\caption{Modulus of the short-time Fourier transform of the sum of 4 vectors of the rotated-time tomogram. 
Each  $u_{m',0}(\theta,j)$ is a bended line in the time-frequency plane where $\theta$ fix the size. 
Here $N=1571$ and $\theta=\pi/8,2\pi/8,3\pi/8,\pi/2$. As $\theta$ increases, the line is stretched to the right 
until it breaks in two parts for $\pi/2$.}\label{rotatomo1}
\label{trot1}
\end{figure}

\begin{figure}
\centering
\includegraphics[width=12cm]{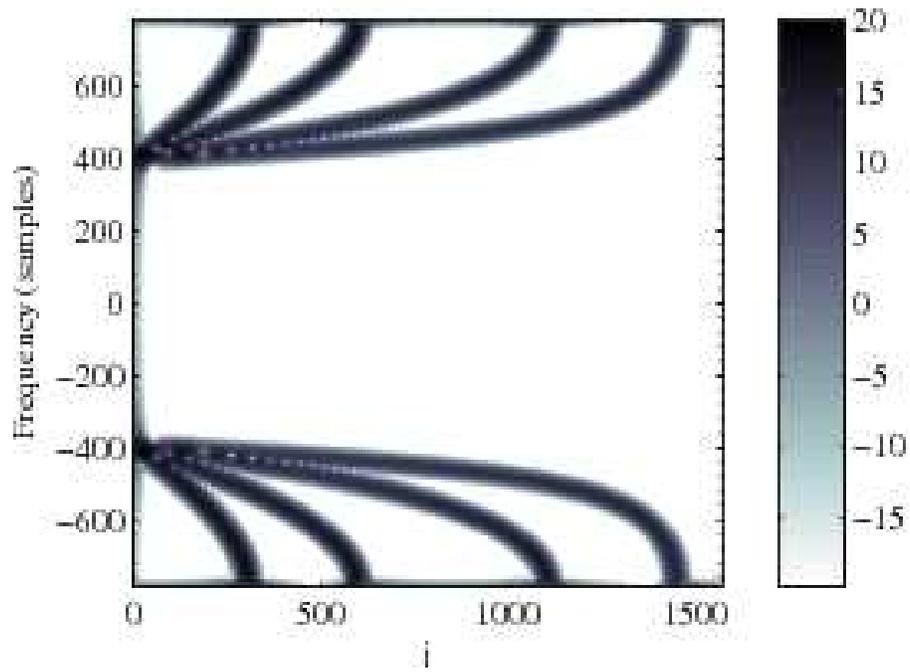}
\caption{Modulus of the short-time Fourier transform of the sum of 4 vectors of the rotated-time tomogram. 
Here $N=1571$ and $\theta=5\pi/8,6\pi/8,7\pi/8$ and $\pi-\pi/16$. Each  $u_{m',0}(\theta,j)$ is made of two bended 
lines in the time-frequency plane, one in the upper-half plane and one in the lower-half plane.}\label{rotatomo2}
\label{trot2}
\end{figure}

How the time-rotated tomogram explores the time-frequency plane is, as
before, illustrated by spectrograms of the eigenstates (Figs.\ref{trot1} and \ref{trot2}). 
For $m=0$, formula (70) reduces to the set of normalized associated Legendre functions $L^{m'}_j$:
$$u_{m',0}(\theta)=\sqrt{\frac{2}{2j+1}}L_j^{m'}(\cos(\theta)).
$$
The normalized associated Legendre functions are related to the unmormalized ones $P^{m'}_j$ through:
$$L_j^{m'}(\cos(\theta))=\sqrt{\frac{2j+1}{2}\frac{(j-m')}{(j+m')}}P^{m'}_j(\cos\theta).
$$
In the tomogram, $\theta$ is the parameter labelling the vectors of the basis associated to $m=0,m'$. 
The index $j$ is the variable. In order to illustrate the effect of this tomogram, we computed numerically 
some vectors in the time-frequency plane (Figs.~\ref{rotatomo1} and~\ref{rotatomo2}).  
In the discrete setting, If we choose $m'=N$, where $N$ is the number of points, the $\{L_j^{N}\}_j$ 
form an orthonormal basis of the discrete time-frequency plane. Hence the projection on the eigenvectors of the 
rotated tomogram with $m=0,m'=N$ can be seen as the projection on the bended lines in the time-frequency plane. 
This tomogram should be adapted for the study of functions which possess certain symetry in the time-frequency plane. 

\section{Hermite basis tomography}

Here we consider a dequantizer 
\begin{equation}
\hat{U}(n,\alpha )=\hat{\mathcal{D}}(\alpha )\mid n\rangle \langle n\mid 
\hat{\mathcal{D}}^{\dagger }(\alpha ),\qquad \alpha =|\alpha |e^{i\theta
_{\alpha }}  \label{PN1}
\end{equation}%
and a quantizer 
\begin{equation}
\hat{D}(n,\alpha )=\frac{4}{\pi (1-\lambda ^{2})}\left( \frac{\lambda +1}{%
\lambda -1}\right) ^{n}\hat{\mathcal{D}}(\alpha )\left( \frac{\lambda -1}{%
\lambda +1}\right) ^{n}\hat{\mathcal{D}}(-\alpha )  \label{PN2}
\end{equation}%
where $-1<\lambda <1$ is an arbitrary parameter and $n$ is related to the
order of an Hermite polynomial. This is analogous to the use of a photon
number basis in quantum optics.

For any signal $f(t)$, one has the probability distribution (tomogram) 
\begin{equation}
\mathcal{M}_{f}(n,\alpha )=\mbox{Tr}\,\mid f\rangle \langle f\mid \hat{U}%
(n,\alpha )  \label{PN3}
\end{equation}%
and, from the tomogram, the signal is reconstructed by 
\begin{equation}
\mid f\rangle \langle f\mid =\sum_{n=0}^{\infty }\int d^{2}\alpha \mathcal{M}%
(n,\alpha )\hat{D}(n,\lambda )  \label{PN4}
\end{equation}%
One has $\mathcal{M}(n,\alpha )\geq 0$ and 
\begin{equation}
\sum_{n=0}^{\infty }\mathcal{M}_{f}(n,\alpha )=1  \label{PN5}
\end{equation}%
for any complex $\alpha $. For an arbitrary operator $\hat{A}$, one has 
\begin{equation}
\hat{I}\hat{A}=\sum_{n=0}^{\infty }\int d^{2}\alpha \hat{D}(n,\alpha )%
\mbox{Tr}\left( \hat{U}(n,\alpha )\hat{A}\right) ,  \label{PN6}
\end{equation}%
where $\hat{I}$ is the identity operator.

The explicit form of the tomogram for a signal function $f(t)$ is 
\begin{equation}
\mathcal{M}_{f}(n,\lambda )=\left\vert \langle f\mid \hat{\mathcal{D}}%
(\alpha )\mid n\rangle \right\vert ^{2}=\left\vert \int f^{\ast
}(t)f_{n,\alpha }(t)\,dt\right\vert ^{2}  \label{PN7}
\end{equation}%
where 
\begin{equation}
f_{n,\alpha }(t)=\hat{\mathcal{D}}(\alpha )\left[ \pi
^{-1/4}(2^{n}n!)^{-1/2}e^{-t^{2}/2}H_{n}(t)\right]  \label{PN8}
\end{equation}%
$H_{n}(t)$ being an Hermite polynomial.

Thus, one has 
\begin{equation}
f_{n,\alpha }(t)=\pi ^{-1/4}(2^{n}n!)^{-1/2}e^{-(\alpha ^{2}-\alpha ^{\ast
2})/4}e^{[(\alpha -\alpha ^{\ast })t]/\sqrt{2}}e^{-\tilde{t}^{2}/2}H_{n}(%
\tilde{t})  \label{PN9}
\end{equation}%
and 
\begin{equation}
\tilde{t}=t-\frac{\alpha +\alpha ^{\ast }}{\sqrt{2}}\,.  \label{PN10}
\end{equation}%
For fixed $|\alpha |$ the tomogram is a function of the discrete set $%
n=0,1,\ldots $ and the phase factor $\theta _{\alpha }$.

\begin{figure}
\centering
\includegraphics[width=12cm]{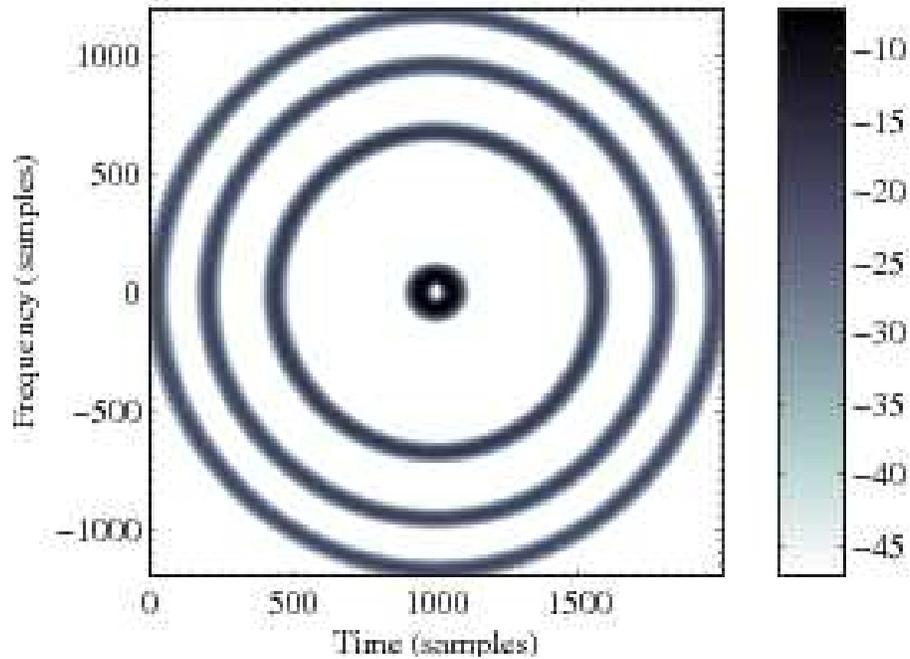}
\caption{Modulus of the short-time Fourier transform of the sum of 4 Hermite functions. Each ring is a Hermite function. Here, the number of points is $N=2000$. The picture has been centered, the origin has been set to Time $t=1000$, Frequency $f=0$. That is to say, $t=-N/2+l\Delta t$ for $l\in[0,N)$, $\Delta t=1$. The smallest circle is for $n=5$ and in increasing size order $n=500$, $n=1000$, $n=1500$, respectively.}\label{Hermite0}
\label{hermite}
\end{figure}

How the Hermite basis tomogram explores the time-frequency plane is, as
before, illustrated by spectrograms of the eigenstates (Fig.\ref{hermite}). 
In the particular case where $\alpha=0$, the functions $f_{n,0}$ are the Hermite functions. 
Their time-frequency representation has been calculated on Figure~\ref{Hermite0}. 
It shows that the tomogram at $\alpha=0$ is suited for rotation invariant functions in the time-frequency plane. 
One can see from (79) that: for real $\alpha$ this pattern is shifted in time and for purely imaginary $\alpha$ 
the pattern is shifted in frequency. The pattern can be shifted in both time and frequency by choosing the 
appropriate complex value for $\alpha$.

\section{Some applications}

The tomograms are squared amplitudes of the signal projections on families
of unitarily equivalent basis (labelled by the $\mu ,\nu $ parameters). By
inspecting the unfolding of these (probability) amplitudes as the parameters
change, several features of the signals are put into evidence. Here we
review briefly three such applications, namely denoising, detection of small
signals and component decomposition, which use the time-frequency tomogram.
Then the time-scale tomogram will be used to analyse a turbulent velocity
fluctuations signal.

For the finite-time signals, instead of (\ref{3.17}), we consider the
finite-time tomogram 
\begin{equation}
M_{1}(\theta ,X)=\left| \int_{t_{0}}^{t_{0}+T}\ f^{*}(t)\psi _{\theta
,X}^{(1)}\left( t\right) \,dt\right| ^{2}=\left| <f,\psi ^{(1)}>\right| ^{2}
\label{4.1}
\end{equation}
with 
\begin{equation}
\psi _{\theta ,X}^{(1)}\left( t\right) =\frac{1}{\sqrt{T}}\exp \left( \frac{%
i\cos \theta }{2\sin \theta }\,t^{2}-\frac{iX}{\sin \theta }\,t\right)
\label{4.2}
\end{equation}
and $\mu =\cos \theta ,\nu =\sin \theta $.

$\theta $ is a parameter that interpolates between the time and the
frequency operators, running from $0$ to $\pi /2$ whereas $X$ is allowed to
be any real number.

\subsection{Detection of small signals}

\begin{figure} [htb]
\centering
\includegraphics[width=12cm]{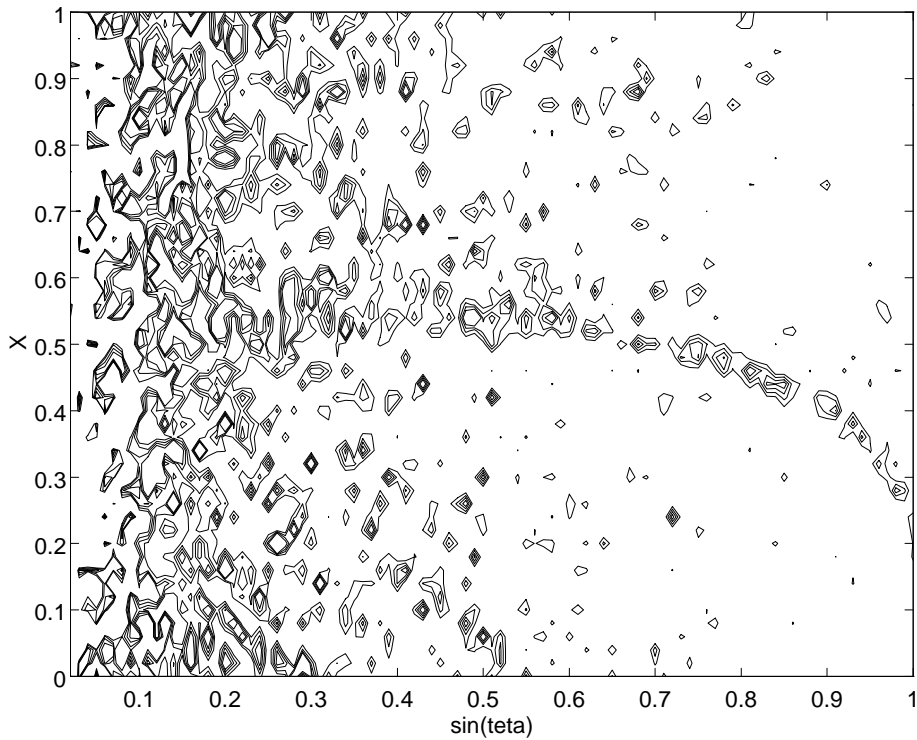} 
\caption{Detection of small signals in noise}
\label{noise}
\end{figure}

As an example \cite{MendesPLA} consider a signal generated as a
superposition of a normally distributed random amplitude - random phase
noise (with total duration $T=1$) with a sinusoidal signal of same average
amplitude, operating only during the time $0.45-0.55$. The signal to noise
power ratio is $1/10$. The true nature of the signal is not revealed neither
from its time development nor from its Fourier spectrum. However computing
the tomogram (see the contour plot in Fig.\ref{noise})
one sees clearly a sequence
of small peaks connecting a time around $0.5$ to a frequency around $200$.
The signature that the signal leaves on the tomogram is a manifestation of
the fact that, despite its low signal to noise ratio, there is a certain
number of directions in the $(t,\omega )$ plane along which detection
happens to be more favorable. For different trials the coherent peaks appear
at different locations, but the overall geometry of the ridge is the same.
On the other hand, a ridge of small peaks is reliable because the rigorous
probability interpretation of $M(\theta ,X)$ renders the method immune to
spurious effects.

\subsection{Denoising and component decomposition}

Most natural and man-made signals are nonstationary and have a
multicomponent structure. Therefore separation of its components is an issue
of great technological relevance. However, the concept of signal component
is not uniquely defined. The notion of \textit{component} depends as much on
the observer as on the observed object. When we speak about a component of a
signal we are in fact referring to a particular feature of the signal that
we want to emphasize. For signals that have distinct features both in the
time and the frequency domain, the time-frequency tomogram is an appropriate
tool.

Here again consider finite-time tomograms as in (\ref{4.1}). For all
different $\theta $'s the $U({\theta })$, of which $B\left( \theta \right) $
is the self-adjoint generator, are unitarily equivalent operators, hence all
the tomograms share the same information.

First we select a subset $X_{n}$ in such a way that the corresponding family 
$\left\{ \psi _{\theta ,X_{n}}^{(1)}\left( t\right) \right\} $ is orthogonal
and normalized, 
\begin{equation}
<\psi _{\theta ,X_{n}}^{(1)}\psi _{\theta ,X_{m}}^{(1)}>=\delta _{m,n}
\label{4.3}
\end{equation}
This is possible by taking the sequence 
\begin{equation}
X_{n}=X_{0}+\frac{2n\pi }{T}\sin \theta \hspace{2cm}n\in \mathbb{Z}
\label{4.4}
\end{equation}
where $X_{0}$ is freely chosen (in general we take $X_{0}=0$). We then
consider the projections of the signal $f(t)$ 
\begin{equation}
c_{X_{n}}^{\theta }(f)=<f,\psi _{\theta ,X_{n}}^{(1)}>  \label{4.5}
\end{equation}

\textit{Denoising} consists in eliminating the $c_{X_{n}}^{\theta }(f)$ such
that 
\begin{equation}
\left| c_{X_{n}}^{\theta }(f)\right| ^{2}\leq \epsilon  \label{4.6}
\end{equation}
for some threshold $\epsilon $. This power selective denoising is more
robust than, for example, frequency filtering which may also eliminate
important signal information.

The \textit{component separation technique} is based on the search for an
intermediate value of ${\theta }$ where a good compromise might be found
between time localization and frequency information. This is achieved by
selecting subsets $\mathcal{F}_{k}$ of the $X_{n}$ and reconstructing
partial signals ($k$-components) by restricting the sum to 
\begin{equation}
f_{k}(t)=\sum_{n\in \mathcal{F}_{k}}c_{X_{n}}^{\theta }(f)\psi _{\theta
,X_{n}}(t)  \label{4.7}
\end{equation}
for each $k$.

As an example consider the following signal

\begin{equation}
y(t)=y_{1}(t)+y_{2}(t)+y_{3}(t)+b(t)  \label{4.8}
\end{equation}
\begin{eqnarray}
y_{1}\left( t\right) &=&\exp \left( i25t\right) ,t\in \left[ 0,20\right] 
\notag \\
y_{2}\left( t\right) &=&\exp \left( i75t\right) ,t\in \left[ 0,5\right] 
\notag \\
y_{3}\left( t\right) &=&\exp \left( i75t\right) ,t\in \left[ 10,20\right]
\label{4.9}
\end{eqnarray}

\begin{figure} [htb]
\centering
\includegraphics[width=12cm]{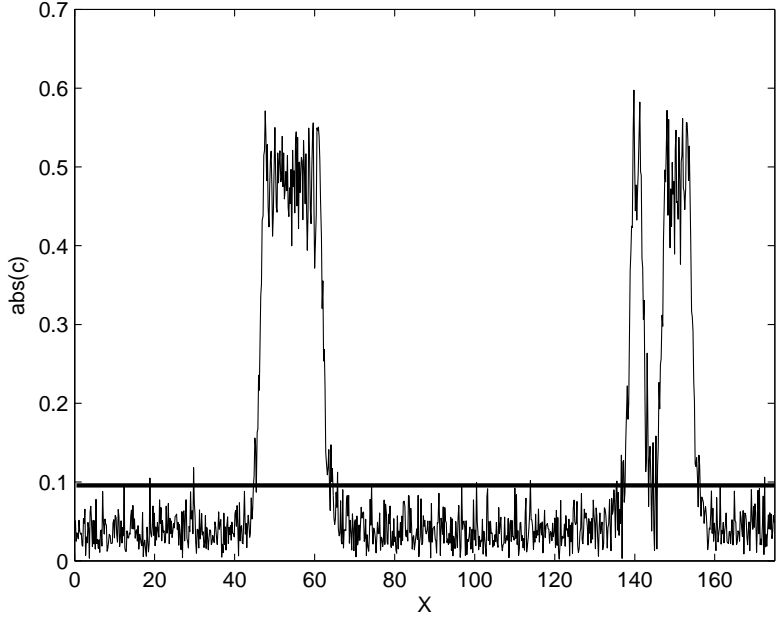} 
\caption{The tomogram at $\protect\theta %
=\protect\pi /5$ used for the component separation}
\label{pi/5}
\end{figure}

\begin{figure} [htb]
\centering
\includegraphics[width=12cm]{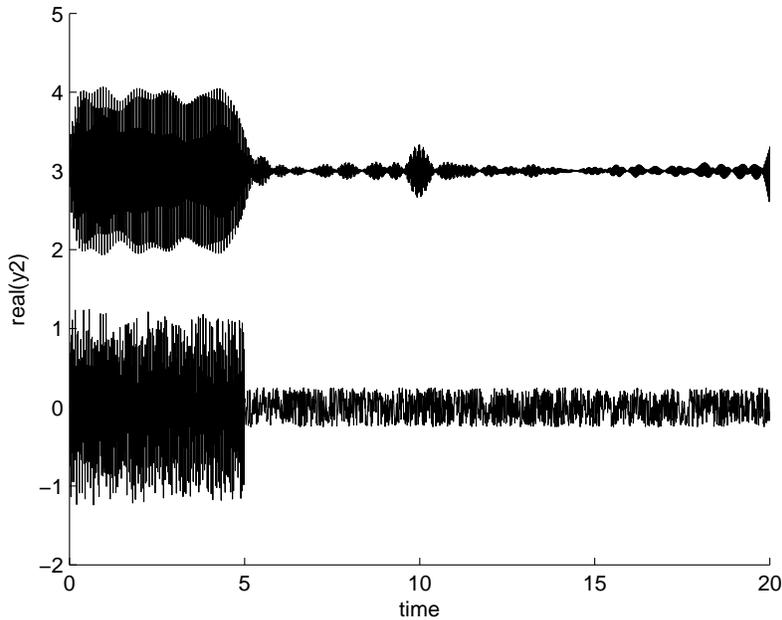} 
\caption{The $y_{2}$ component}
\label{y2component}
\end{figure}

\begin{figure} [htb]
\centering
\includegraphics[width=12cm]{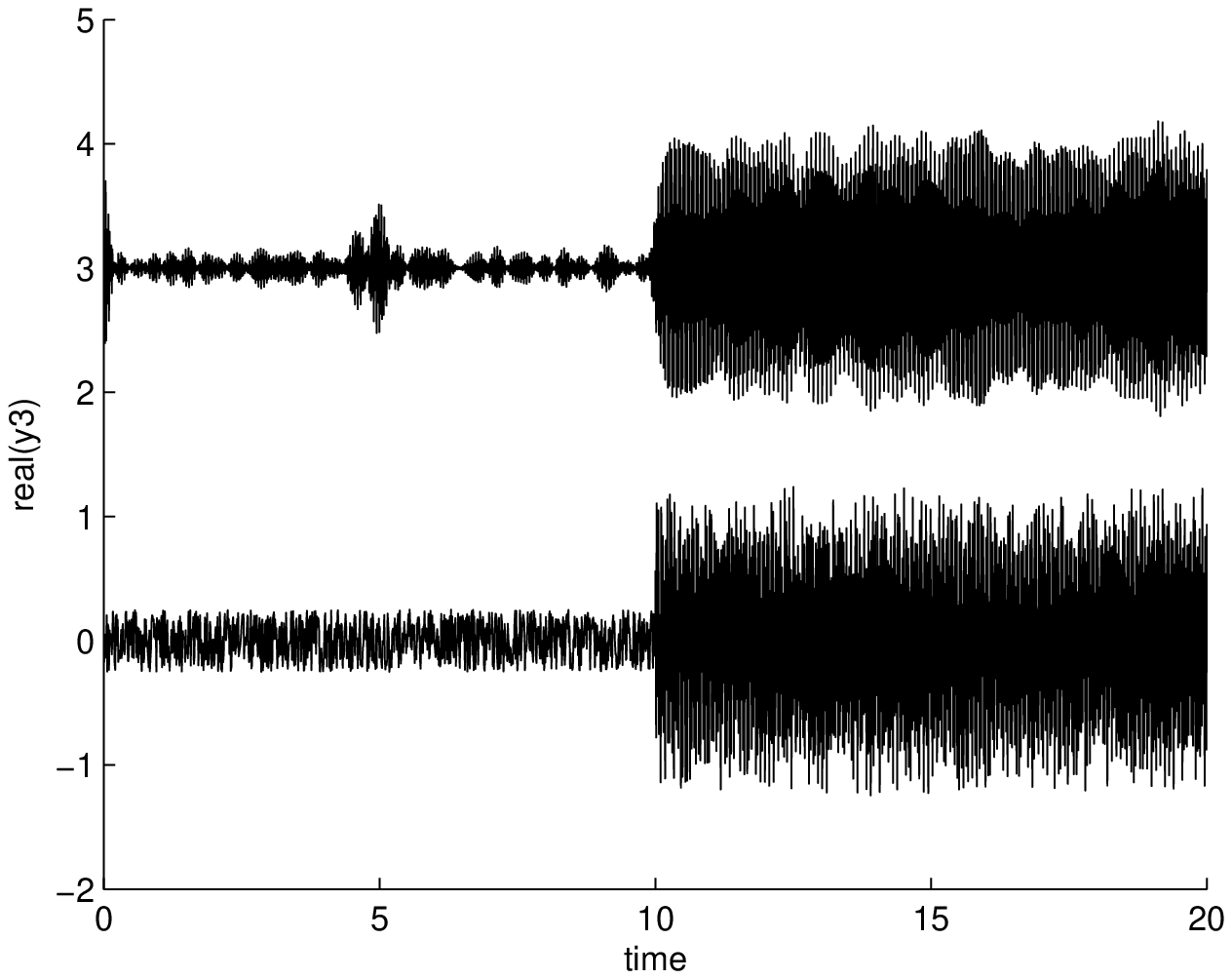} 
\caption{The $y_{3}$ component}
\label{y3component}
\end{figure}

Separation is impossible both at the time ($\theta =0$) and the frequency ($%
\theta =\frac{\pi }{2}$) axis. However, at some intermediate $\theta $ value
one obtains distinct probability peaks (Fig.\ref{pi/5}), which after the projections (%
\ref{4.7}) allows an accurate separation of the signal components (Figs.\ref{y2component}
and \ref{y3component})

Component decomposition of more complex signals (nonlinear chirps
overlapping in both the time and the frequency domains and experimental
reflectometry signals) has been successfully carried out by this technique 
\cite{reflecto1} \cite{reflecto2}.

\subsection{Tomograms and turbulent velocity fluctuations}

Here we report briefly on an analysis by the tomographic technique of a
velocity fluctuation signal of a turbulent flow in a wind tunnel. It
illustrates the fact that the choice of the pair of non-commuting operators
in tomogram, should be adapted to the signal under study. As before we use
finite-time tomograms in the interval $\left( t_{0},t_{0}+T\right) $. For
the finite-time (time-frequency) tomogram $M_{1}$, the normalization and a
set of $X_{n}$'s leading to an orthonormalized set of eigenstates has
already been written in (\ref{4.1})-(\ref{4.2}).

For future reference we include here the corresponding sets of
orthonormalized eigenstates for the finite-time time-scale tomogram $%
M_{2}(\mu ,\nu ,X)$ (Eq.\ref{3.18}) and for the finite-time time-conformal
tomogram $M_{4}(\mu ,\nu ,X)$ (Eq.\ref{3.20}): 
\begin{equation}
M_{2}\left( \theta ,X\right) =\left\vert \int_{t_{0}}^{t_{0}+T}\ f^{\ast
}(t)\psi _{\theta ,X}^{(2)}\left( t\right) \,dt\right\vert ^{2}=\left\vert
<f,\psi ^{(2)}>\right\vert ^{2}  \label{4.10}
\end{equation}%
\begin{equation}
\psi _{\theta ,X}^{(2)}\left( t\right) =\frac{1}{\sqrt{\log \left\vert
t_{0}+T\right\vert -\log \left\vert t_{0}\right\vert }}\frac{1}{\sqrt{%
\left\vert t\right\vert }}\exp i\left( \frac{\cos \theta }{\sin \theta }\,t-%
\frac{X}{\sin \theta }\,\log \left\vert t\right\vert \right)  \label{4.11}
\end{equation}%
\begin{equation}
X_{n}=X_{0}+\frac{2n\pi }{\log \left\vert t_{0}+T\right\vert -\log
\left\vert t_{0}\right\vert }\sin \theta \hspace{2cm}n\in \mathbb{Z}
\label{4.12}
\end{equation}%
and 
\begin{equation}
M_{4}(\theta ,X)=\left\vert \int_{t_{0}}^{t_{0}+T}\ f^{\ast }(t)\psi
_{\theta ,X}^{(4)}\left( t\right) \,dt\right\vert ^{2}=\left\vert <f,\psi
^{(4)}>\right\vert ^{2}  \label{4.13}
\end{equation}%
\begin{equation}
\psi _{\theta ,X}^{(4)}\left( t\right) =\sqrt{\frac{t_{0}\left(
t_{0}+T\right) }{T}}\frac{1}{\left\vert t\right\vert }\exp i\left( \frac{%
\cos \theta }{\sin \theta }\,\,\log \left\vert t\right\vert +\frac{X}{t\sin
\theta }\right)  \label{4.14}
\end{equation}%
\begin{equation}
X_{n}=X_{0}+\frac{t_{0}\left( t_{0}+T\right) }{T}2\pi n\sin \theta \hspace{%
2cm}n\in \mathbb{Z}  \label{4.15}
\end{equation}

\begin{figure} [htb]
\centering
\includegraphics[width=12cm]{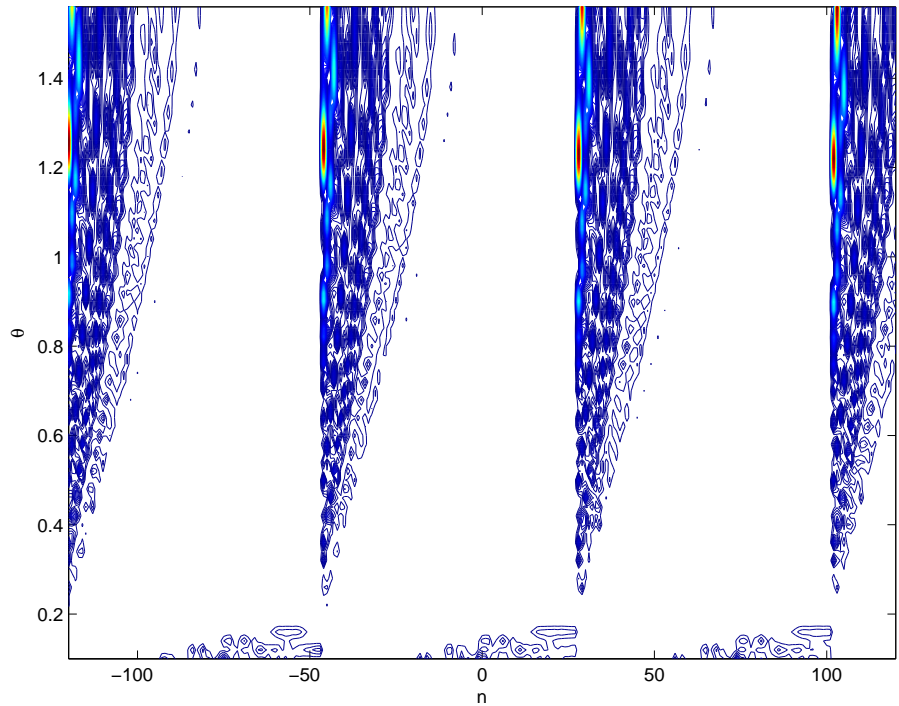} 
\caption{Contour plot of the tomogram
for a velocity fluctuations signal}
\label{velocity1}
\end{figure}

\begin{figure} [htb]
\centering
\includegraphics[width=12cm]{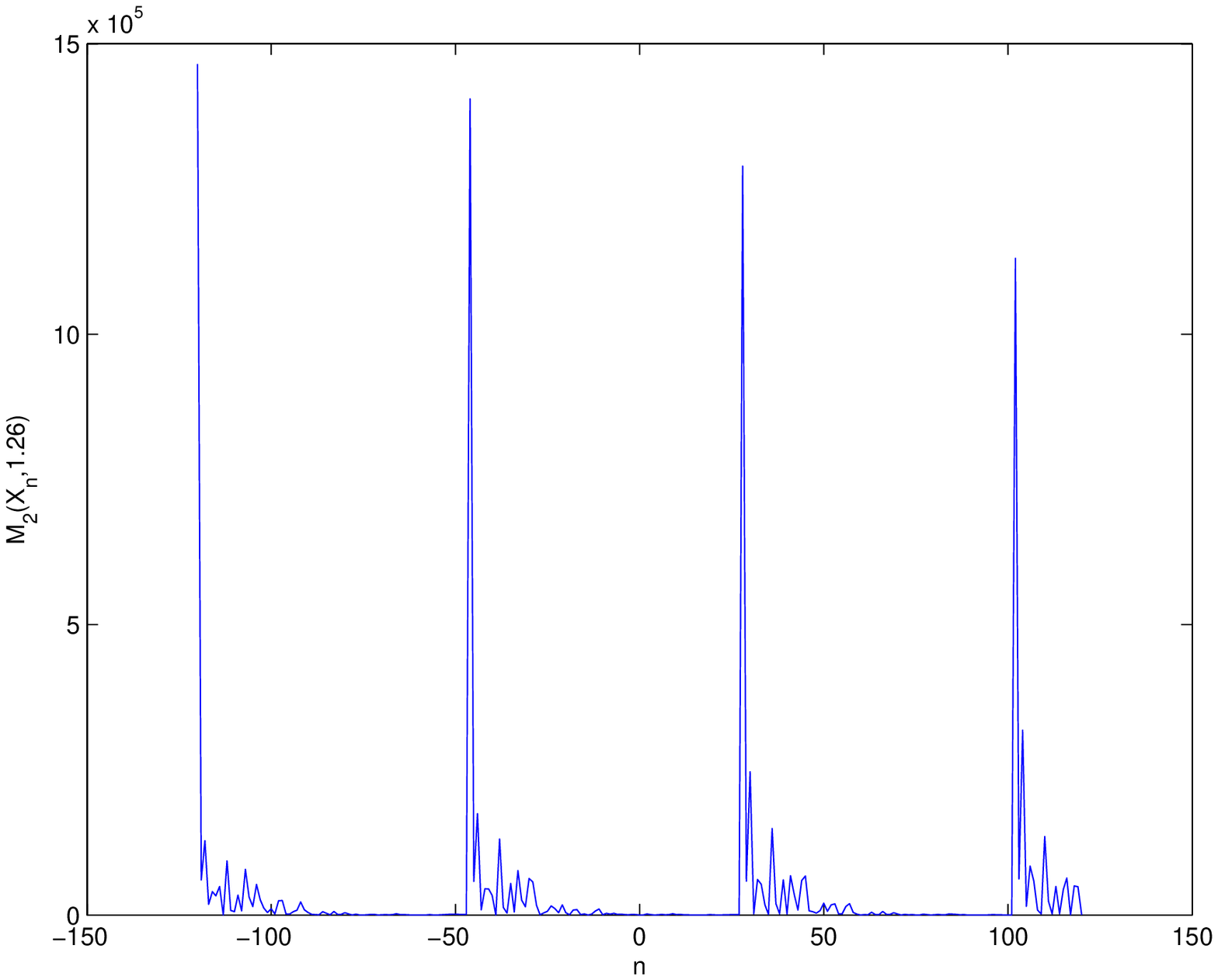} 
\caption{The tomogram $M_{2}\left( X_{n},%
\protect\theta \right) $ at $\protect\theta =1.26$}
\label{velocity2}
\end{figure}

\begin{figure} [htb]
\centering
\includegraphics[width=12cm]{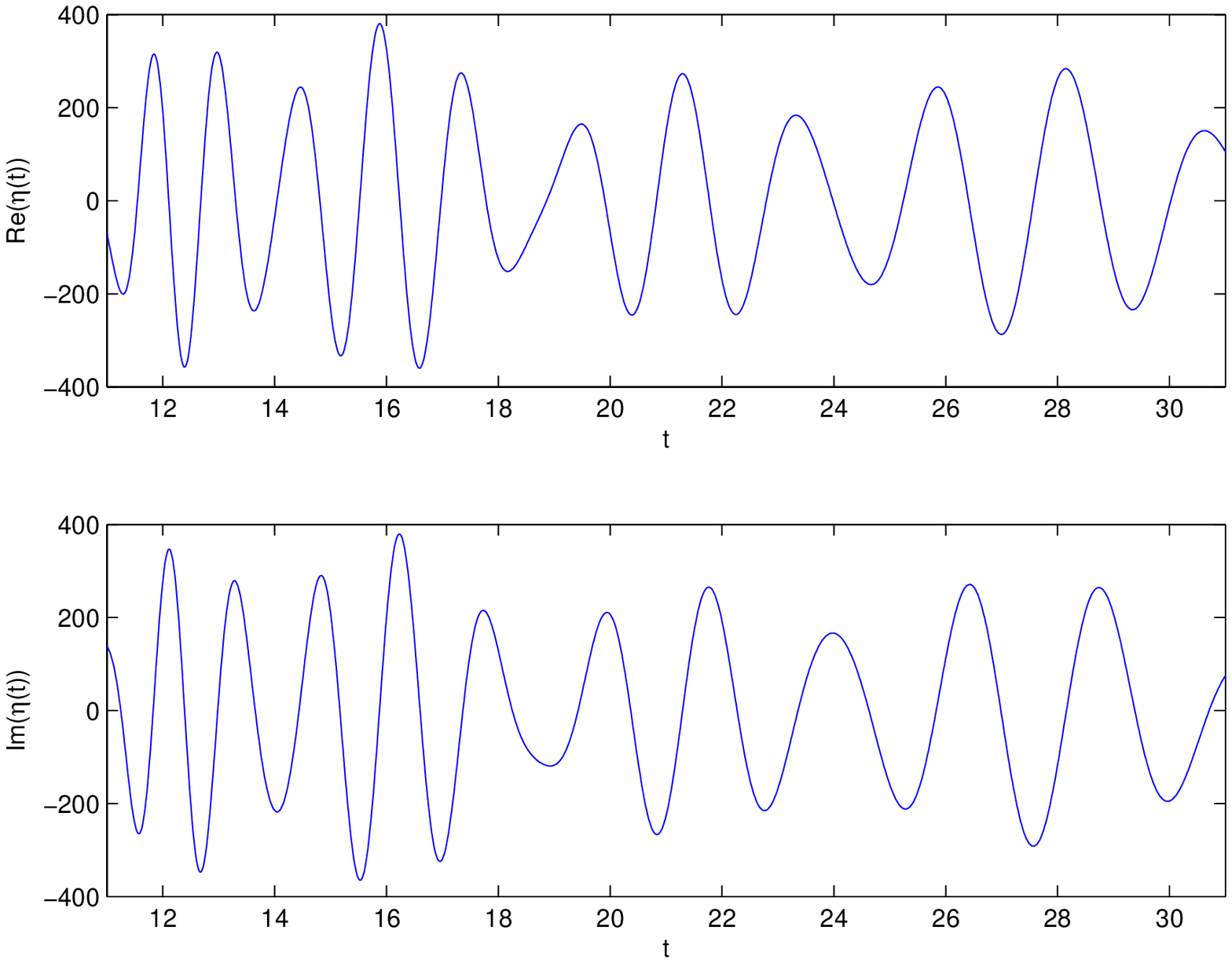} 
\caption{The structure $\protect\eta %
\left( t\right) $ corresponding to the second peak in Fig.7}
\label{structure}
\end{figure}

Analyzing the turbulent velocity fluctuations signal with these tomograms,
one notices that except for some features on the frequency axis
corresponding to some dominating frequencies, no interesting structures are
put into evidence when one use the time-frequency tomogram. The situation is
more interesting for the time-scale tomogram $M_{2}\left( \theta ,X\right) $%
. In Fig.\ref{velocity1} we show a contour plot for $M_{2}\left( \theta ,X\right) $
corresponding to a section of 1000 data points. For intermediate regions of $%
\theta $ one notices, a strong concentration of energy in a few regions.
This is put into evidence by a cut at $\theta =1.26$ (Fig.\ref{velocity2}). Projecting out
the signal corresponding to these regions with the corresponding $\psi
_{\theta ,X}^{(2)}\left( t\right) $'s at this $\theta $, one sees that
although the signal has many complex features most of the energy is
concentrated in fairly regular structures. Fig.\ref{structure} shows the structure $\eta
\left( t\right) $ corresponding to the second peak in Fig.\ref{velocity2}.

\section{Conclusions}

Tomograms provide a two-variable characterization of signals which, due to
its rigorous probabilistic interpretation, is robust and free of artifacts
and ambiguities. For each particular signal that one wants to analyse the
choice of the appropriate tomogram depends not only on the signal but also
on the features that we might want to identity or emphasize. So far we have
explored component separation, denoising and identification \ of small
signal in noise, but other features may also benefit from the robust
probabilistic of the tomographic analysis. This was our main motivation to
include here a long list of many different operator choice leading to
different classes of tomograms.

The description of the tomograms as operator symbols, with the corresponding
quantizers and dequantizers, not only provides an alternative formulation
but may also be used to extend the algebraic signal processing formalism to
a wider nonlinear context.

\end{document}